\documentstyle[12pt]{article}

        {\newpage
        \setcounter{page}{1}}

\renewcommand{\title}[1]{%
  {\begin{center} \Large\bf #1
        \end{center}}
      \vskip .3in}

\renewcommand{\author}[1]{%
  {\begin{center} #1
        \end{center}}}

\renewcommand{\abstract}[1]{%
        \begin{center}%
        {\vspace{1em}\vspace{0pt}\bf Abstract}%
        \end{center}%
        \noindent #1}

\renewcommand{\date}[1]{%
        \begin{center}%
        #1%
        \end{center}}

      {\end{thebibliography}}

\makeatletter
        \@addtoreset{equation}{section}%
\makeatother

\newcommand{\eqn}[1]{\label{eq:#1}}
\newcommand{\refeq}[1]{(\ref{eq:#1})}
\newcommand{\refeqs}[2]{(\ref{eq:#1}-\ref{eq:#2})}
 
 \newcommand{\Eqs}{Eqs.~\refeqs}

\newcommand{\beq}{\begin{eqnarray}}
\newcommand{\eeq}{\end{eqnarray}}

\newcommand{\naive}{{na\"\i ve}} 
\newcommand{\Naive}{{Na\"\i ve}} 
\newcommand{\smt}{$SU(3)\times SU(2)\times U(1)$}

\newcommand{\drawsquare}[2]{\hbox{%
\rule{#2pt}{#1pt}\hskip-#2pt
\rule{#1pt}{#2pt}\hskip-#1pt
\rule[#1pt]{#1pt}{#2pt}}\rule[#1pt]{#2pt}{#2pt}\hskip-#2pt
\rule{#2pt}{#1pt}}
\newcommand{\Yasymm}{\raisebox{-3.5pt}{\drawsquare{6.5}{0.4}}\hskip-6.9pt%
        \raisebox{3pt}{\drawsquare{6.5}{0.4}}}
      
\newcommand{\mybar}[1]%
{\kern 0.8pt\overline{\kern -0.8pt#1\kern -0.8pt}\kern 0.8pt}
\newcommand{\sla}[1]%
        {\raise.15ex\hbox{$/$}\kern-.57em #1}
\newcommand{\roughly}[1]%
{\mathrel{\raise.3ex\hbox{$#1$\kern-.75em\lower1ex\hbox{$\sim$}}}}

\newcommand{\jref}[4]{{ #1} {\bf #2}, #3 (#4)}

\newcommand{\PLB}[3]{\jref{Phys.\ Lett.}{#1B}{#2}{#3}}

\newcommand{\PRD}[3]{\jref{Phys.\ Rev.}{D#1}{#2}{#3}}

\newcommand{\PRL}[3]{\jref{Phys.\ Rev.\ Lett.}{#1}{#2}{#3}}

 \def\CO{{\cal O}}
\def\CL{{\cal L}}   \def\CW{{\cal
    W}}   
\def\ltap{\ \raise.3ex\hbox{$<$\kern-.75em\lower1ex\hbox{$\sim$}}\ }
\def\gtap{\ \raise.3ex\hbox{$>$\kern-.75em\lower1ex\hbox{$\sim$}}\ }

\begin{document}

\begin{titlepage}
\begin{center}
  {\hbox to\hsize{ \hfill BUHEP-97-16 }}  {\hbox
    to\hsize{hep-th/9706275 \hfill DOE/ER/40561-328-INT97-00-172}}  {\hbox
    to\hsize{ \hfill UW/PT-97-13}}

  \bigskip \bigskip \bigskip \vskip.2in

  {\Large \bf Counting $4\pi$'s in  Strongly Coupled Supersymmetry} \\

  \bigskip \bigskip \bigskip \vskip.2in

  {\bf Andrew G. Cohen$^a$, David B. Kaplan$^b$ and  Ann E. Nelson$^c$}\\

  \vskip.2in {\small \sl $^a$ Department of Physics, Boston University
    Boston, MA 02215, USA }

  {\tt cohen@andy.bu.edu}

  \bigskip\bigskip

  { \small \sl $^b$ Institute for Nuclear Theory, Box 351550

    University of Washington, Seattle, WA 98195-1550 }

  {\tt dbkaplan@phys.washington.edu}

  \bigskip \bigskip

  { \small \sl $^c$ Department of Physics, Box 351560 University of
    Washington, Seattle, WA 98195-1560 }

  {\tt anelson@phys.washington.edu}

  \bigskip \bigskip

  \vspace{1.5cm}
  {\bf Abstract}\\
\end{center}

\bigskip

We extend the ``\naive\ dimensional analysis'' arguments used in QCD for
estimating the strengths of operators in chiral Lagrangians to
strongly coupled supersymmetric theories.  In particular, we show how
to count factors of $4\pi$---an inexact science, but nevertheless a
useful art when such theories are used to model real particle physics.
\bigskip

\end{titlepage}

\newpage
\section{Introduction}

Recently there has been a surge of interest in constructing models of
new physics in terms of strongly coupled
supersymmetry (SUSY).  These strong interactions typically
produce dynamical SUSY breaking, composite quarks and leptons, or both
\cite{SUSY,comp}. The low energy descriptions of such models inevitably
involve an effective field theory, an expansion in local operators
with unknown coefficients. Discussions of the phenomenology of such
models require estimates of the sizes of the operator coefficients
which control parameters of direct experimental interest, such as
squark and gaugino masses, CKM mixing angles, {\it etc}. Usually we
are thwarted in this endeavor by our ignorance of underlying strong dynamics;
nevertheless estimates can be made on the basis of dimensional
analysis and $4\pi$ counting. In refs. \cite{comp,effsusy}, we
exploited a dimensional analysis scheme generalized from QCD; in this
note we make our analysis explicit---the analysis itself is model
independent.

The sizable mass gap between the pions and
all other hadronic states in QCD leads to a profitable analysis of low
energy hadronic physics in terms of an effective field theory, the
chiral Lagrangian.  Like all effective field theories, the chiral
Lagrangian is constructed as an expansion in local operators
constrained by low energy symmetries; each operator is multiplied by a
characteristic mass scale to an appropriate power, times a
dimensionless coefficient.  In order to estimate the effect of
operators neglected in a calculation, it is useful to have a method
for estimating the sizes of these dimensionless coefficients.  Such a
method was introduced by Weinberg \cite{weinberg} and discussed in
detail in ref. \cite{georgi}.  The method is predicated on the
assumption that an effective field theory of strongly interacting
fields has operator coefficients such that radiative corrections are
no larger than $\CO(1)$ times tree level coefficients.  Assuming
that the radiative corrections are in fact of the same size as
tree level coefficients leads to ``\naive\ dimensional analysis''
(NDA) estimates for the size of interactions.

We begin by explaining the power counting arguments for conventional
field theories in a manner which differs somewhat from that in the literature,
using the language of the Wilsonian renormalization group.  We
also discuss operator matching and the inclusion of
light, weakly coupled fields. We then turn to the supersymmetric
generalization. In our conclusions we discuss some of the assumptions
behind NDA estimates.

\subsection{\Naive\ dimensional analysis}

We begin by assuming that we have some strongly interacting theory
that we would like to match onto an effective theory at a scale
$\Lambda$.  The effective action which describes the interactions of
any massless scalar fields $\Phi$ and fermion fields $\Psi$ in a
derivative expansion is assumed to be characterized by a single
dimensionful scale\footnote{This assumption
could conceivably
be wrong, {\it e.g.} for
  strongly coupled theories which are near an infrared fixed point,
  such as ``walking technicolor''\cite{walking}. Also, we are
  neglecting the possibility of small dimensionless
  numbers such as $1/N_c$.}.
  In a nonsupersymmetric theory, the $\Phi$ fields
will be massless only if they are Goldstone bosons, while the fermions
can be protected from acquiring mass by chiral symmetry.  The
effective theory is described in terms of a Wilsonian effective action
at the scale $\Lambda$:
\beq
\eqn{gaction} S_\Lambda = {1\over g^2}
\int {\rm d}^4x\ \Lambda^4\ \hat \CL_{\Lambda}\left({\Phi'\over
    \Lambda}, {\Psi'\over
    \Lambda^{3/2}},{\partial\over\Lambda}\right)\ ,
\eeq
where $g$ is
a dimensionless parameter which we will determine.  Terms in
$\hat\CL_\Lambda$ are assumed to have $\CO(1)$ coefficients, and the
scale $\Lambda$ is the matching scale between the UV and IR
descriptions of the theory, or equivalently, the mass scale of degrees
of freedom that have been integrated out.  (The primes on the fields
$\Phi'$, $\Psi'$ are a reminder that kinetic terms may not have a
canonical normalization in this basis.)  Upon integrating out modes in
the momentum shell $[e^{-1}\Lambda,\Lambda]$, the operator
coefficients in $\CL$ will flow to new values.  Contributions to the
operator coefficients of the effective action at $L$ loops will be of
characteristic size
\beq
{1\over g^2} \left({g^2\over
    16\pi^2}\right)^L
\eeq
(see ref. \cite{coleman} for example) where
we have included a factor of $1/16\pi^2$ for each loop. We will have
a ``natural'' theory if these renormalizations are no larger than tree
level, which requires $ g\ltap 4\pi$; the NDA assumption is that this
inequality is saturated,
\beq
\eqn{nda} g\sim 4\pi\ .
\eeq
 If $g$ were to be
smaller, we would assign it to weak rather
than strong coupling.
We will
assume that eq.~\refeq{nda} holds throughout this paper, and
examine this assumption in our conclusions.

We may rescale the fields to recast the effective action
\refeq{gaction} into a form with conventionally normalized kinetic
terms\footnote{We are concerned only with factors of coupling
  constants and $4\pi$'s; consequently this normalization may differ
  from a truly conventional one by factors of order 1.}:
\beq
\Phi' =  g \Phi,\qquad \Psi' = g \Psi
\eeq
so that the
effective action \refeq{gaction} becomes
\beq
\eqn{gactionii}
S_\Lambda = \int {\rm d}^4x\ { \Lambda^4 \over g^2}\ \hat
\CL_{\Lambda}\left({g\Phi\over \Lambda},{ g\Psi\over
    \Lambda^{3/2}},{\partial\over \Lambda}\right)\ .
\eeq

The above results may be compared with discussions of power counting
in the chiral Lagrangian by using the correspondence
\beq
\Lambda \to
\Lambda_{\chi}\ ,\qquad {\Lambda\over g} \to f_\pi
\eeq
where $\Lambda_{\chi}$ is
called the chiral symmetry breaking scale, and $f_\pi$ is the pion
decay constant.  Eq. \refeq{gactionii} now reads
\beq
\eqn{gactionchi}
S = \int {\rm d}^4x\ { \Lambda_{\chi}^2 f_\pi^2}\ \hat
\CL_{\Lambda}\left({\pi^a\over f_\pi}, {\Psi\over \sqrt{\Lambda_{\chi}
      f_\pi^2}},{\partial\over \Lambda_{\chi}}\right)\ ,
\eeq
where $\pi^a$
are the pions; $\Psi$ might be light fermions such as chiral quarks,
or heavy nucleons for which there are sources.  The result
\refeq{gactionchi} agrees with the results of ref. \cite{georgi}.

\subsection{Weakly coupled light fields}

Besides the pions, the low energy theory may contain light, weakly
interacting fields, such as the photon. For the photon we are guided
by gauge invariance, and make the replacement $\partial \to (\partial
- i e A)$ in the action \refeq{gactionii}:
\beq
\eqn{gactionA}
S_\Lambda = \int {\rm d}^4x\ { \Lambda^4 \over g^2}\ \hat
\CL_{\Lambda}\left({\Phi'\over \Lambda},{ \Psi'\over \Lambda^{3/2}},{e
    A\over \Lambda},{\partial\over \Lambda}\right) + \CL_w(A)\ .
\eeq
Note that we must include an independent Lagrangian to account for
interactions among the weakly coupled fields ({\it eg\/}, the kinetic
term for the photons).
For a weakly coupled field, the factor of $g\simeq 4\pi$ is
replaced by a perturbative coupling, $e$ in this case. This suggests a
more natural form for (\refeq{gactionA}):
\beq
\eqn{gactionA2}
S_\Lambda = \int {\rm d}^4x\ { \Lambda^4 \over g^2}\left[\hat
\CL_{\Lambda}\left({\Phi'\over \Lambda},{ \Psi'\over
    \Lambda^{3/2}},{\hat e 
    A'\over \Lambda},{\partial\over \Lambda}\right) +
\frac{1}{{\hat e}^2}\hat\CL_w\left({{\hat e} A' \over \Lambda}\right)\right]\ .
\eeq
where $A'=gA$ and $\hat e \equiv e/g \simeq e/(4\pi)$. Weak coupling now
corresponds to $\hat e << 1$.

While the form of photon interactions is dictated by gauge
invariance, the power counting is clearly the same for any weakly
coupled field.  While strongly coupled fields appear in the
combination $g\Phi/\Lambda \simeq 4\pi \Phi/\Lambda$, a weakly coupled
field $\phi$ appears as ${\hat e} g \phi/\Lambda$, that is with an extra factor
of the weak coupling $\hat e$. This procedure works for
nonrenormalizable interactions as well: a weakly interacting field
which couples to $m$ strongly interacting 
fields via a dimension $4+n$
operator with  coefficient $(4\pi)^{(m-1)}/M^n$  also appears as
${\hat e} g \phi/\Lambda$ in the effective theory, with a
dimensionless weak coupling of \beq\eqn{nonren}{\hat e} \to\Lambda^n/M^n\ .\eeq

\subsection{Matching operators}
In QCD one needs to match operators involving quarks and gluons ({\it e.g,}
four quark operators from the weak interactions, $GG$,
$GG\tilde G$, {\it etc.}) onto operators in the effective theory.  In order
to estimate the size of these operators in the effective theory, we
continue in the spirit of \naive\ dimensional analysis and assume that
extra loops in a diagram at the matching scale do not change the
characteristic size of an amplitude in both the UV
and the IR descriptions of the theory, meaning that the strongly
coupled particles in either description couple with strength $g\sim
4\pi$.  The computation is simplest in the primed normalization of Eq.
\refeq{gaction}---operators constructed out of quark and gluon
fields $q'$, $G'$ match onto operators with the same symmetry
properties constructed out of composite fields $\Phi'$, $\Psi'$ with
dimensions matched by powers of $\Lambda$, and no dimensionless
coefficients other than $\CO(1)$.  For example, consider how various
QCD operators constructed of quarks and gluons map onto operators in
the chiral Lagrangian constructed of pions $\pi'$ or nucleons
$N'$:
\beq
\bar q' q'
&\to& c_1 \Lambda  \pi'\pi' + c_2   \bar N' N' + \cdots\ ,\cr
(\bar u_L^\prime\gamma^\mu d_L^\prime)(\bar d_L^\prime\gamma_\mu
u_L^\prime) &\to&  c_3\Lambda^2\partial_\mu\pi'\partial^\mu\pi' + c_4\Lambda^3
\bar N'
N' + \cdots \,\cr G'G' &\to&  c_5\partial\pi'\partial\pi'  +
c_6\Lambda \bar N' N' + \cdots\ .
\eeq
where the dots represent all other operators  consistent with the
symmetries, and the $c$'s are dimensionless numbers of order one.
To express this mapping in terms of conventionally
normalized fields, we need only rescale all of the fields by a power
of $g\simeq 4\pi$:
\beq
\bar q q &\to& c_1 \Lambda  \pi\pi + c_2  \bar
N N\ ,\cr (\bar u_L\gamma^\mu d_L)(\bar d_L\gamma_\mu
u_L) &\to&  c_3 {\Lambda^2\over 16\pi^2}
\partial_\mu\pi\partial^\mu\pi
+ c_4{\Lambda^3
  \over 16 \pi^2}{\bar N} N + \cdots \,\cr
GG &\to& c_5 \partial\pi\partial\pi + c_6 \Lambda \bar N N + \cdots\ .
\eeq

In summary, matching an operator with
$n$ strongly interacting fields in the UV to operators
with $m$ composite fields in the IR entails the appropriate power of
$\Lambda$ to match
the dimensions and
an explicit factor of $(4\pi)^{n-m}$.
It should be noted that the $\Delta I=1/2$ rule is a notorious
failure of such power counting arguments, since some of the NDA estimates for
the matching of weak four quark operators are off  by a factor of
$\sim 10$,  except when analyzed within
the context of the chiral quark model \cite{halfi}.

\section{\Naive\ dimensional analysis for supersymmetric theories}
The above analysis carries over to supersymmetric theories with little
modification.  The main difference is that one needs to extend the
power counting scheme to include auxiliary $F$ and $D$ fields.  The
supersymmetric generalization of eq. \refeq{gaction} is (ignoring
gauge interactions for now)
\beq
\eqn{susyi} S_\Lambda = {1\over g^2}
\int {\rm d}^4x \ \Lambda^4 \left[ \int {\rm d}^2 \theta\,{\rm
    d}^2\bar\theta\ \Lambda ^{-2} \hat K + \int {\rm d}^2 \theta\
  \Lambda ^{-1}\ \hat W+ \int {\rm d}^2\bar\theta\ \Lambda ^{-1}\ \hat
  W^*\right]
\eeq
where the dimensionless K\"ahler potential $\hat K$
and superpotential $\hat W$ are functions constructed out of the
dimensionless superfields
\beq
{1\over\Lambda}
\Phi'(x,\theta/\sqrt{\Lambda}) = {1\over\Lambda} \left( A' + {\theta
    \Psi'\over\sqrt{\Lambda}} + {\theta^2 F'\over \Lambda}\right)\ ,
\eeq
and supersymmetric derivative
\beq
{D\over \Lambda} \sim
{\partial_\theta + i \bar\theta\sigma\cdot \partial_x \over \Lambda}\
,
\eeq
with coefficients of $\CO(1)$.  To regain a canonical
normalization, we make the substitution
\beq
\eqn{susyii}
{1\over \Lambda} \Phi'(x,\theta/\sqrt{\Lambda}) = {g\over \Lambda}
\Phi(x,\theta/\sqrt{\Lambda})\ ,
\eeq
with $g\simeq 4\pi$.

As in the non-SUSY case, weakly coupled chiral superfields $\phi$
interact with composite superfields in the combination
\beq
{\hat\lambda}{g\over \Lambda} \phi(x,\theta/\sqrt{\Lambda})\ ,
\eeq
where
$\hat \lambda$ is the weak coupling.  Canonically normalized weak gauge
superfields $V$ and spinor superfields $\CW_\alpha$ couple as
\beq
{\hat e} {g\over \Lambda} V(x,\theta/\sqrt{\Lambda})\qquad{\rm
  and}\qquad {{\hat e} g\over
  \Lambda^{3/2}} \CW_\alpha(x,\theta/\sqrt{\Lambda})\ ,
\eeq
where $e=\hat e g$ is a perturbative gauge coupling.  For composite
gauge superfields, such as occur in a free magnetic phase \cite{FMP},
the factor $\hat e$ is of order one at the scale $\Lambda$.

\section{Examples}
We give two examples of the power counting described above.  The first
example is an asymptotically free supersymmetric gauge theory of the
``$s$-confining'' type  discussed in
\cite{sconfine}. The
second example is the Effective SUSY theory introduced in
\cite{effsusy}, which involves analysis of an effective action for
which the UV description is unknown.

\subsection{A model with composites and a calculable superpotential}

The first example we consider is a theory which in the UV is an
$Sp(4)$ gauge theory with a single antisymmetric tensor field $A$ and six
fundamental fields $Q$ \cite{sp2n}; it has much in common with the
phenomenological models discussed in \cite{comp}.  The theory confines,
and the composites relevant for the IR description are
(with non-canonical normalization):
\beq
\eqn{composites}
 T_2^\prime={1\over 8\Lambda} (A'A')=1_{-6,0}\ ,\quad
 M_0^\prime=\frac{(Q'Q')}{\Lambda}={\Yasymm\,}_{2,2/3}\ ,\quad
 M_1^\prime=\frac{(Q'A'Q')}{\Lambda^2}={\Yasymm\,}_{-1,2/3}
\eeq
where we have listed the quantum
numbers of the moduli under the $SU(6)\times U(1)\times U(1)_R$ global
symmetry of the model. The dynamically generated superpotential is
\beq
\eqn{wdyni} W_{dyn} \propto \left( {1\over 3\Lambda} T_2^\prime
  M_0^{\prime 3} + {1\over 2}  M_0^\prime  M_1^{\prime 2}\right)\ .
\eeq
The relative factor
of the operators in \refeq{wdyni} is fixed by requiring the correct
constraints on the moduli.

\subsubsection{The effective strong interactions}

To rewrite $W_{dyn}$ in terms of canonically normalized fields, we
perform the operator matching as in \S1.3
\beq
\eqn{opmap} T_2={1\over
  8} (AA)\left({ 4\pi\over \Lambda}\right)\ ,\qquad M_0=(QQ)\left({
    4\pi\over \Lambda}\right) ,\qquad M_1=(QAQ)\left({ 4\pi\over
    \Lambda}\right)^2\ ,
\eeq
Following the NDA prescription of
\Eqs{susyi}{susyii}, the dynamically generated superpotential
\refeq{wdyni} takes the form
\beq
\eqn{wdynii} W_{dyn} \simeq 4\pi
\left( {4 \pi\over 3\Lambda } T_2 M_0^3 + {1\over 2} M_0 M_1^2\right)\ .
\eeq
In particular, note that the dynamically generated Yukawa
coupling among the $M_0$ and $M_1$ component fields is
of order $4\pi$.

The relative factor between the two terms, reflecting the constraints
on the moduli, is preserved by this rescaling. However note that the
kinetic terms for these fields (from the K\"ahler potential) may still
contain unknown order one coefficients.

K\"ahler potential terms must be consistent with the global
symmetries of the theory; this includes terms such as
\beq
K&=& \sum_i a_i
\vert \Phi_i\vert^2 + \left({1\over \Lambda}\right)^2 \sum_{i} c_{i}
\Phi_i^* D^2 \Phi_i + \left({4\pi\over \Lambda}\right)^2 \sum_{ij}
c_{ij} \vert \Phi_i\vert^2 \vert \Phi_j\vert^2 \cr &+&
\left({4\pi\over \Lambda}\right)^3 \tilde c T_2 M_0^2 (M_1^*)^2 +
{h.c.} +\ldots
\eeq
where $\Phi_i=\{T_2,M_0,M_1\}$, and the $a_i,c$,
$\tilde c$ coefficients are $\CO(1)$.  Note that while each new field
brings with it a factor of $4\pi/\Lambda$, the momentum expansion is
in powers of $p/\Lambda$.  This is consistent with having integrated
out heavy fields with masses $M=\Lambda$ and couplings $g=4\pi$.

\subsubsection{Perturbative interactions and spurion analysis}

We now consider how to construct the effective theory when
a perturbative superpotential in the UV description of the theory is
included.
Following the
discussion of \Eqs{susyi}{susyii} we include  superpotential
perturbations of the form
\beq
\eqn{wpertii} W_{pert} = \epsilon {1\over
  g^2} \Lambda^3 \hat W(\Phi_i^\prime/\Lambda),
\eeq
with $g=4\pi$. Weak coupling then corresponds to $\epsilon \ll 1$.
As an example we may take
\beq
\eqn{wperti} W_{pert} =\frac{1}{16\pi^2}
\left[ \epsilon_1 \phi' (A'A') +
\epsilon_2 (Q'A'Q')+ {\epsilon_3
  \over \Lambda^2} (Q'Q')(Q'A'Q')\right]
\eeq
where $\phi'$ is a superfield which is neutral
under the strong $Sp(4)$ gauge group, and for simplicity, we have
suppressed $SU(6)$ indices.
In terms of more conventionally normalized fields this is
\beq
\eqn{wpertiii} W_{pert} =
{4\pi\epsilon_1} \phi (AA) + {4\pi\epsilon_2} (QAQ)+
\epsilon_3 {(4\pi)^3 \over \Lambda^2} (QQ)(QAQ).
\eeq
The perturbative parameters $\epsilon_i$
may be treated as spurions, each carrying $SU(6)\times U(1)\times
U(1)_R$ quantum numbers
\beq
\epsilon_1 \phi= 1_{6,2}\ ,\qquad
\epsilon_2={ \mybar \Yasymm\,}_{1,4/3}\ ,\qquad {\epsilon_3} ={
  \Yasymm\, }_{-1,2/3}\ .
\eeq
The quantum numbers of the spurions,
along with holomorphy, constrain the induced superpotential to
consist of only three terms. Using the operator mapping
\refeq{opmap} we find
\beq
W^{eff}_{pert} \approx  a_1 \epsilon_1 \Lambda \phi T_2 +
a_2 {\epsilon_2 \over
  4\pi}\Lambda^2 M_1 + a_3 \epsilon_3\Lambda M_0M_1 \ ,
\eeq
while the K\"ahler potential
will include such terms as
\beq
K_{pert} = \sum_{ij} a_{ij} \vert
\Phi_i\vert^2 \vert \epsilon_j\vert^2 + \left({4\pi\over
    \Lambda}\right)^2\sum_i b_i \vert \Phi_i\vert^2 \vert \epsilon_1
\phi\vert^2 + \ldots
\eeq
where $a,b$ are coefficients of order one.

Note that the superfields receive
perturbative corrections to the leading term in the K\"ahler
potential.  We must avoid an $\epsilon$ dependent rescaling of the
fields, however, if we wish to maintain holomorphy in the
superpotential.

\subsection{Scalar and Gaugino masses in strongly coupled theories with
  Supersymmetry Breaking}

We now use our $4\pi$ counting scheme to analyze the superpartner
masses in theories with a strongly coupled sector in which
supersymmetry is broken. In order to systematically discuss
supersymmetry breaking effects, we assume that the supersymmetry
breaking scale in the low energy effective theory is below the scale
$\Lambda$, so that supersymmetry is linearly realized in the low
energy effective theory; this allows a weakly coupled description of
supersymmetry breaking in terms of a nonzero vacuum expectation value
for the $F$-term of some ``composite'' superfield $\Phi$.  Such
theories have been considered in refs.~\cite{SUSY, effsusy, others}.
We assume the mechanism for ordinary gaugino masses involves conventional
\smt\ gauge interactions, and that these interactions are weakly
coupled at the matching scale.  A generic theory where strongly
interacting particles  carrying \smt\ interactions are integrated
out at the scale $\Lambda$ may induce terms in the low energy effective theory
of the form
\beq\eqn{gaugino}
\int d^2\theta \left[c {\hat e}^2
\frac{(\Phi')^n}{
 \Lambda^n}
\CW_\alpha\CW^\alpha\right]_F\ ,
\eeq
where $\hat e$ is an \smt\ gauge coupling $e$ divided by $4\pi$, $\CW$ is
an \smt\ gauge spinor superfield, $\Phi'=4\pi \Phi$, and we
expect the coefficient $c$ to be of order one. If we have
the correct degrees of freedom to describe the true ground state, then
$\langle\Phi'\rangle < \Lambda$ for all $\Phi$. Thus the maximum size
for the light gaugino masses $\tilde m_i$ occurs
when there is a term of the form
eq.~\refeq{gaugino} with $n=1$ for a composite
field $\Phi$ with a nonzero $F$-term,
so that at the scale $\Lambda$
\beq
\tilde m_{\rm gaugino}=
{4\pi c}\frac{e^2}{16\pi^2}
\frac{ F_{\Phi}}{\Lambda}\ .
\eeq

For the squark and slepton masses, two options have been
considered. The first ``gauge mediated'' \cite{GMSB} solution has the
squarks and sleptons communicate with the supersymmetry breaking
sector only via ordinary gauge interactions. Then the squark and
slepton masses squared will arise from terms 
which are induced only by graphs which involve at least one weak gauge
loop (and an arbitrary number of strong loops). Integrating
out strongly interacting fields results in couplings of quarks and
leptons to the light composite fields: in general such loop effects
can only
appear in the K\"ahler potential and will be proportional to  factors
of  ${\hat e}^4 \phi^{'\dagger} \phi'/\Lambda^2$ where $\phi$ is
a quark or lepton superfield, and $\phi'=4\pi\phi$. 
Squark and slepton masses squared may be obtained from the induced operator
\beq
\eqn{sfermion}\int d^4\theta c'\frac{{\hat e}^4}{16\pi^2}
\frac{\Phi'^\dagger\Phi'}{\Lambda^2} \phi'^\dagger \phi'= 
\int d^4\theta c'\left(\frac{e^2}{16\pi^2}\right)^2
\frac{16\pi^2\Phi^\dagger\Phi}{\Lambda^2} \phi^\dagger \phi\ ,
\eeq
where $c'$ is or order one,
leading to squark and slepton masses squared of order
\beq\tilde m_{\rm sfermion}^2=c'
\left({4\pi }\frac{e^2}{16\pi^2}
\frac{ F_{\Phi}}{\Lambda}\right)^2\ .\eeq
 Thus the phenomenologically
desirable relation that squark and gluino masses are comparable in
magnitude is obtained.

Alternatively, one could follow the ``effective supersymmetry'' 
approach of ref.~\cite{effsusy}
and allow some of the quark and lepton superfields to be
composites of strongly interacting fields\footnote{Note that the
  quantity which we called $\Lambda$ in ref.~\cite{effsusy} is
  equivalent to what we are calling $\Lambda/g=\Lambda/(4\pi)$ in this
  work.}.
The natural size of the composite squark and slepton masses is then
much larger than that of the \smt\ gauginos, and is of order
\beq\tilde m_{\rm comp\ sfermion}^2\sim
\left({4\pi }
\frac{ F_{\Phi}}{\Lambda}\right)^2\,,\eeq
which corresponds to (\refeq{sfermion}) with $\hat e \to 1$.
Such a theory with phenomenologically acceptable gaugino masses,
$m_{\rm gaugino}\gtap 100$~GeV, cannot
give rise to natural electroweak symmetry breaking unless at least
one Higgs superfield and its ``brothers'' \cite{dg} ({\it i.e}  those
superfields with $\CO(1)$ couplings to the Higgs), are elementary,
weakly coupled fields. It is however possible in a natural theory to
have the first two generations of quarks and leptons carry the
strong supersymmetry breaking interactions. The corresponding squarks
and sleptons will have masses larger than those of the other
superpartners by a factor of $\sim 16\pi^2/e^2$~\cite{effsusy}.

\section{Conclusions}
The simple arguments presented here allow the construction of a natural
effective field theory with only minimal understanding of the
underlying dynamics, provided the strong dynamics of the
underlying theory is characterized
by a single scale. We have given simple power counting rules
for the  factors of $4\pi$ in the coefficients of
dynamically generated superpotentials and K\"ahler potentials in a
strongly coupled supersymmetric theory. We have also given a simple
algorithm for
counting the $4\pi$ factors in operators involving both weakly coupled
superfields and
composite superfields, and in symmetry breaking ``spurion'' factors.

All these applications may be summarized in a simple and general
principle, embodied in eqs. \refeq{gactionA}\ and \refeq{susyi}:
when the theory is characterized by a single dimensionful parameter
$\Lambda$, all dimensionless operator coefficients are naturally of order $1$
in a normalization in which the effective action has an overall factor
of $1/g^2\approx 1/(16 \pi^2)$. Provided this same normalization is used, weak
couplings or spurions (that is,  operators with coefficients
parametrically smaller than $1/(16 \pi^2)$) may be directly included
according to the same rule, with the same parametric factor in the UV
and IR. Matrix elements of operators follow the same rule: operators
with no explicit factors of $4\pi$ match onto operators also with no
explicit factors of $4\pi$.

What if the strongly coupled theory is characterized by more than one
scale? If these scales are widely separated, a sequence of effective
field theories following the above rules may still be constructed,
starting with the highest scale first. When these scales are not
widely separated, or when the strong dynamics is spread out over a
large energy range (such as near an approximate IR fixed point) NDA
is likely to fail.

Why is the assumption $g \sim 4 \pi$ reasonable?
Generically we expect that, given a weakly
coupled theory in the UV, as we scale down in energy irrelevant
operators become weaker, while the effective couplings of relevant
operators become stronger. When these couplings become of order $4
\pi$, the theory is strongly coupled and will typically undergo a
phase transition; we then construct the effective action in terms of
new degrees of freedom, which will again be weakly coupled, with some
relevant and some irrelevant operators. As we scale down in energy the
process continues, until we reach a phase in which there are no
relevant operators, and hence the theory remains weakly coupled. For
example, in QCD the derivative coupling of the pions means all
interactions are irrelevant, and the pions are weakly coupled at low
energies. For the supersymmetric examples presented here the quartic
and Yukawa couplings, marginal at tree level, become irrelevant when
radiative corrections are included.

The simple power counting rules given here have not considered
factors that may be associated with numbers of fields, for example the
number of flavors $N_f$, the number of preons in a composite superfield,  
or the number of colors $N_c$---inclusion of such
factors is straightforward, and should improve the accuracy of NDA.

As this paper was being completed,  ref.~\cite{luty}
appeared which deals with
similar issues.

\section{Acknowledgments}
A.G.C. is supported in part by the DOE under grant
\#DE-FG02-91ER40676; D.B.K is supported in part by DOE grant
DOE-ER-40561, and NSF Presidential Young Investigator award
PHY-9057135. A.E.N. was supported in part by the
DOE under grant \#DE-FG03-96ER40956.

\end{document}